# Stability and Performance Analysis for SISO Incremental Flight Control

Zhidong Lu  Florian Holzapfel

*Institute of Flight System Dynamics, Technische Universität München, Garching, 85748, Germany*

***ABSTRACT***

**Incremental Nonlinear Dynamic Inversion (INDI) control has attracted increasing research attention for it retains the high-performance of NDI and has enhanced robustness. However, when actual elements of the flight control system and real-world phenomena (such as actuator dynamics, sensor noise, time delay, etc.) are considered, the INDI control may have degraded performance or even lose stability. This paper analyzed the stability and performance of an incremental controller for a SISO linear plant based on transfer functions. Besides, the theoretical analysis results are verified through numerical evaluations for the incremental controller of the short-period aircraft model using comprehensive metrics.**

# 1 Introduction

Nonlinear Dynamic Inversion (NDI) is a widely used multivariable control technique in the aerospace industry [1]. The major advantage of NDI is avoiding gain-scheduling by directly canceling nonlinearities with feedback linearization. But it requires accurate knowledge of the nonlinear dynamics and is thus sensitive to model uncertainty [2]. Incremental Nonlinear Dynamic Inversion (INDI) is a variation of NDI with better robustness because it uses the (angular) acceleration feedback to reduce model dependency [3]. Successful applications of INDI flight control have been conducted on MAVs and multi-rotors platforms [4]-[6], where good robustness and disturbance rejection were shown. Stability and robustness analysis for ideal INDI control has been conducted in Ref. [7].

However, there are still challenges for practical applications of INDI. For example, the time delay can cause significant performance degradation of the flight control system [8], and INDI is very susceptible to sensor time delay [9]. Other issues, including actuator dynamics [10], input delay [11], noise filtering [12], and model uncertainties [13] also pose challenges. Besides, the strategy of Pseudo Control Hedging (PCH) was first developed for approximate dynamic inversion based model reference adaptive control to deal with input dynamics[14], and it has also been directly transplanted to INDI control[15][16]. However, the application of PCH to INDI still needs further investigation. Therefore, it is of interest to know about the stability and performance of incremental flight control under practical circumstances.

In this paper, we discussed the stability and performance of the incremental controller for Single-input-single-output (SISO) LTI flight dynamics using classical control theories. Elements of the flight control system, including actuator, sensor, filter, error control gain, and some real-world phenomena like time delay and model mismatch, are taken into account. This paper is structured as follows: Sec. 2 derives control law and analytical transfer functions of the incremental controller. Sec. 3 and Sec. 4 address the stability and performance issues. Numerical evaluations are conducted for a short-period model incremental controller in Sec. 5. Finally, Sec. 6 gives conclusions and recommendations.

# 2 Incremental Control Law and Transfer Functions

Consider the generic form of a SISO linear system,

$$\begin{cases} \dot{x} = Ax + Bu \\ y = Cx \end{cases} \tag{1}$$

where $x \in \mathbb{R}^n$ is the state vector, $u$ is the input, $y$ is the controlled variable. Two forms of incremental control law, the conventional incremental dynamic inversion, and the modified incremental dynamic inversion, will be discussed below.

## 2.1 Conventional incremental dynamic inversion

The conventional incremental dynamic inversion method is based on the computation of the control increment at the current time step with respect to the system's condition in the past time step.

$$\dot{x} = \dot{x}_0 + A(x - x_0) + B(u - u_0) = \dot{x}_0 + A \cdot \Delta x + B \cdot \Delta u \tag{2}$$

It is assumed that the state changes significantly slower than the control input in a minimal time increment. By assuming $\Delta x = 0$, Equation (2) can be simplified as:

$$\dot{x} = \dot{x}_0 + B \cdot \Delta u \tag{3}$$

Combining Equation (3) with (1), we have:

$$\dot{y} = \dot{y}_0 + CB \cdot \Delta u \tag{4}$$

Inverting Equation (4) and setting a pseudo control $\nu = \dot{y}$, the following control law is obtained:

$$\Delta u = (CB)^{-1}(\nu - \dot{y}_0) \tag{5}$$

The total input will be the sum of the current input $u_0$ and the calculated input increment $\Delta u$:

$$u = u_0 + \Delta u \tag{6}$$

In practical applications, either the input measurement value $u_{0mea}$ or the estimation $u_{0mdl}$ from an actuator model will be used as $u_0$.

## 2.2 Modified incremental dynamic inversion

The conventional incremental control law builds a direct connection between the commanded pseudo control and the command input increment:

$$\Delta\nu_c = CB \cdot \Delta u_c \tag{7}$$

However, this design regards the expected bandwidth of the pseudo control variable and the actuator bandwidth as consistent, but the two are likely and usually expected to be inconsistent. The expected bandwidth of the pseudo control variable could be controlled by:

$$\dot{\nu}_c = K_\nu \Delta\nu_c \tag{8}$$

and the achieved bandwidth of a first-order actuator model is determined by its time constant:

$$\dot{u}_c = \frac{1}{T_{act}}(u_c - u) = \frac{1}{T_{act}}\Delta u_c \tag{9}$$

Based on Eq. (7) and assume the time increment is infinitely small, we have:

$$\dot{\nu}_c = CB \cdot \dot{u}_c \tag{10}$$

Combine Eq.(8)~(10), one obtains the modified incremental control input as:

$$\Delta u_c = T_{act}(CB)^{-1} K_\nu \Delta\nu_c \tag{11}$$

And the final control input will be:

$$u_c = u_0 + \Delta u_c \tag{12}$$

It can be seen that the conventional incremental control law is a particular case of the modified incremental control law when $T_{act}K_\nu = 1$. Therefore, the modified control law will provide an additional degree of freedom for control system tuning.

With the following first-order command reference model, the Pseudo Control Hedging (PCH) strategy is used [15][16].

$$\dot{r} = K_r(c - r) \tag{13}$$

The hedging signal is designed as the difference between the commanded pseudo control and the achieved pseudo control:

$$\nu_h = \nu_c - \nu_a = CB(u_c - \hat{u}) \tag{14}$$

The hedging signal is then introduced into the reference model to modify the state response, while the instantaneous pseudo-control output of the reference model remains unchanged:

$$\dot{r} = \nu_r - \nu_h, \nu_r = K_r(c - r) \tag{15}$$

The block diagram illustrates the modified incremental control law is given by Fig. 1. Other elements of the incremental controller are also shown, including:

1) The error controller $K_P$, which is a scalar value;
2) The filter to estimate the angular acceleration from measurements is simplified as a combination of a first-order low pass filter $F(s) = 1/(T_{\mathrm{diff}}s + 1)$ and the differential operator $s$ (This form of representation is for the convenience of subsequent analysis);
3) The control allocation module $\hat{B}^{-1} = (CB)^{-1}$, which is a scalar value for a SISO plant;

4) The actuator dynamics, modeled as a first-order transfer function with time delay $G_a(s) = \frac{e^{-\tau_a s}}{(T_{\text{act}}s+1)}$;

5) The sensor dynamics $H$, modeled as a transfer function with time delay $H(s) = \frac{e^{-\tau_S s}}{(T_{\text{sensor}}s+1)}$;

6) The actuator measurements dynamics $G_{am}$.

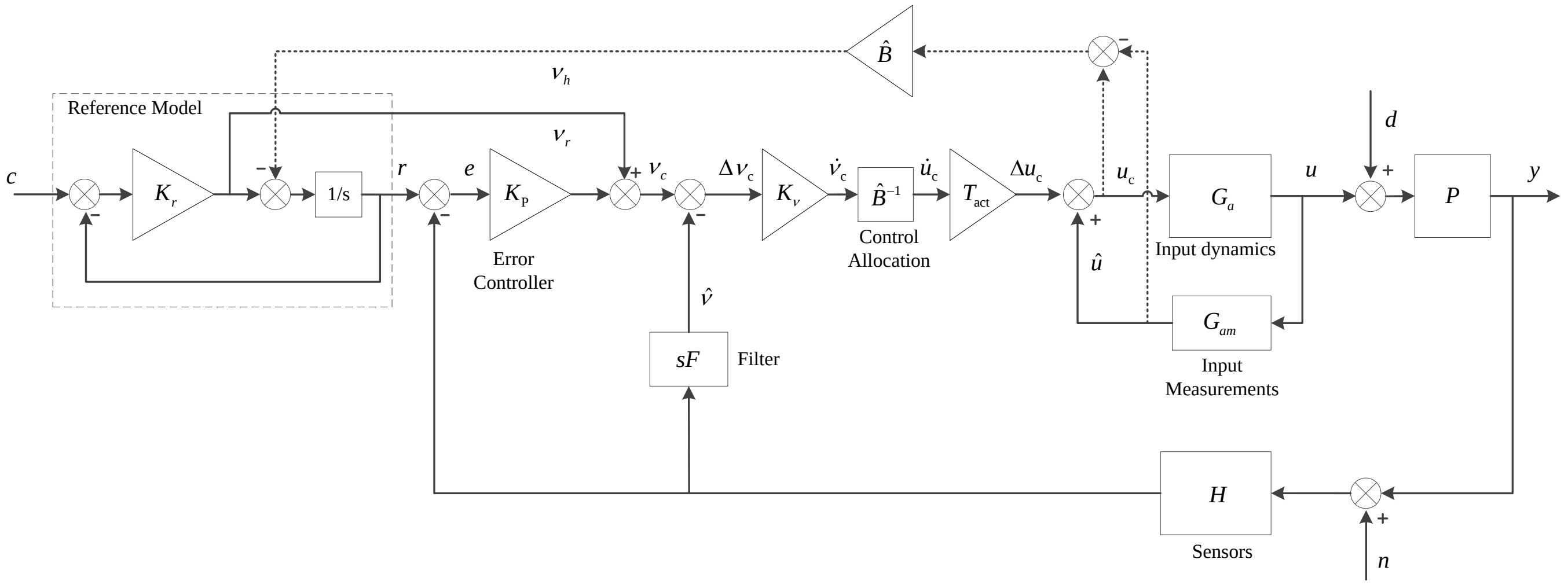

**Fig. 1 Block diagram of the SISO incremental control system.**

## 2.3 Transfer functions

The open-loop and closed-loop transfer functions for both the conventional and modified incremental controller are derived in this section.

*A. Conventional incremental controller*

When PCH is off, the open-loop transfer function broken at input *u* is:

$$L_u = G_a(s)\bar{C}(s)H(s)P(s) = G_a(s)\frac{\big(K_P + sF(s)\big)}{\hat{B}[1 - G_a(s)G_{am}(s)]}H(s)P(s) \tag{16}$$

where $\bar{C}(s)$ is an equivalent INDI controller. And the closed-loop transfer function from command to roll rate output is:

$$T_{yc} = \frac{G_a(s)P(s)(K_P + s)}{\left(\frac{1}{K_r}s + 1\right)\left[\hat{B}\big(1 - G_a(s)G_{am}(s)\big) + G_a(s)H(s)P(s)\big(K_P + sF(s)\big)\right]} \tag{17}$$

When PCH is on, the open-loop transfer function is given by:

$$L_{u,PCH} = G_a(s)\bar{C}_{\text{PCH}}(s)H(s)P(s) = G_a(s)\frac{(s + K_r)\big(K_P + sF(s)\big)}{\hat{B}(K_P + s)\big(1 - G_{am}(s)G_a(s)\big)}H(s)P(s) \tag{18}$$

And the closed-loop transfer function from command to output is:

$$T_{yc,\mathrm{PCH}} = \frac{G_a(s)P(s)(K_P+s)K_r}{\hat{B}\left[(K_P+s)\left(1-G_a(s)G_{am}(s)\right)\right]+G_a(s)H(s)P(s)(K_r+s)\left(K_P+sF(s)\right)} \quad (19)$$

*B. Modified incremental controller*

Based on Fig.1, one can find that the modified incremental controller can be derived from the conventional one by extending the control allocation module in the following way:

$$\hat{B}^{-1} \Leftarrow \hat{B}^{-1}T_{act}K_v \quad (20)$$

Similarly, the open-loop and closed-loop transfer functions can be obtained:

$$L_u = G_a(s)\bar{C}(s)H(s)P(s) = G_a(s)\frac{T_{act}K_v\left(K_P+sF(s)\right)}{\hat{B}[1-G_a(s)G_{am}(s)]}H(s)P(s) \quad (21)$$

$$T_{yc} = \frac{T_{act}K_vG_a(s)P(s)(K_P+s)}{\left(\frac{1}{K_r}s+1\right)\left[\hat{B}\left(1-G_a(s)G_{am}(s)\right)+T_{act}K_vG_a(s)H(s)P(s)\left(K_P+sF(s)\right)\right]} \quad (22)$$

$$L_{u,PCH} = G_a(s)\bar{C}_{\mathrm{PCH}}(s)H(s)P(s) = \frac{G_a(s)T_{act}K_v(s+K_r)\left(K_P+sF(s)\right)H(s)P(s)}{\hat{B}[K_PK_vT_{act}+s+K_r(1-K_vT_{act})][1-G_a(s)G_{am}(s)]} \quad (23)$$

$$T_{yc,\mathrm{PCH}} = \frac{T_{act}K_vG_a(s)P(s)(K_P+s)K_r}{\hat{B}[K_PK_vT_{act}+s+K_r(1-K_vT_{act})][1-G_a(s)G_{am}(s)]+G_a(s)T_{act}K_v(s+K_r)(K_P+sF(s))H(s)P(s)} \quad (24)$$

Only the modified incremental controller will be analyzed in the following.

# 3 Stability Analysis for Incremental Controller

This section analyzes the incremental controller's stability, especially the influence of time delay, control gain, actuator dynamics, sensor dynamics, filter, and Pseudo-Control-Hedging on stability.

## 3.1 Ideal case

Ideally, assuming $H(s)=1, F(s)=1, G_{am}(s)=1$, then the equivalent incremental controller becomes:

$$\bar{C}(s) = \frac{T_{act}K_v(K_P+s)(T_{act}s+1)}{\hat{B}T_{act}s} = \frac{(K_PT_{act}+1)K_v}{\hat{B}} + \frac{T_{act}K_v}{\hat{B}}s + \frac{K_vK_P}{\hat{B}}\frac{1}{s} \quad (25)$$

In this condition, the incremental controller is reduced to a PID controller, and the proportional, integral, and derivative coefficients are respectively $\bar{C}_P = \frac{(K_PT_{act}+1)K_v}{\hat{B}}, \bar{C}_I = \frac{K_vK_P}{\hat{B}}, and\ \bar{C}_D = \frac{T_{act}K_v}{\hat{B}}$ . When $K_P, K_v$ and $T_{act}$ increase or $\hat{B}$ decreases, all three coefficients will increase.

The open-loop transfer function is given by:

$$L_u = G_a(s)\bar{C}(s)H(s)P(s) = \frac{K_v(K_P+s)}{\hat{B}s}P(s) \quad (26)$$

We can find that the actuator dynamics are canceled in the open-loop transfer function. And the incremental controller is reduced to a PI controller for the plant $P(s)$. The open-loop frequency response goes:

$$|L_u(j\omega)| = \frac{K_v\sqrt{\omega^2 + K_P^2}}{\hat{B}\omega}|P(j\omega)|, \qquad \angle L_u(j\omega) = \angle P(j\omega) + arctan\frac{\omega}{K_P} - \frac{\pi}{2} \tag{27}$$

It can be inferred that a larger $K_P$ leads to increased magnitude and decreased phase; thus the gain margin, phase margin, and time delay margin are all reduced; $K_v$ and $\hat{B}$ only affect the magnitude and a larger $K_v$ or smaller $\hat{B}$ will increase the magnitude and result in reduced gain margin and time delay margin.

**Example 1**. Take the aircraft roll mode dynamics for example, which is $\dot{p} = L_p p + L_{\delta_a}\delta_a$. The stability margins of the incremental controller for the roll mode dynamics can be derived as follows:

$$GM = \infty, PM = \arctan\frac{\omega_c}{K_P} - \arctan\frac{L_p}{\omega_c}, TDM = \frac{1}{\omega_c}\left(\arctan\frac{\omega_c}{K_P} - \arctan\frac{L_p}{\omega_c}\right) \tag{28}$$

where $\omega_c$ is the magnitude cross-frequency:

$$\omega_c = \sqrt{\frac{K_v^2 - \left(L_p\frac{\widehat{L_{\delta_a}}}{L_{\delta_a}}\right)^2 + \sqrt{\left(K_v^2 - \left(L_p\frac{\widehat{L_{\delta_a}}}{L_{\delta_a}}\right)^2\right)^2 + 4\left(\frac{\widehat{L_{\delta_a}}}{L_{\delta_a}}\right)^2 {K_P}^2 K_v^2}}{2}} \tag{29}$$

Assume $\widehat{L_{\delta_a}} = L_{\delta_a}$, the time delay margin of the roll mode incremental controller with different values of $[K_P \quad K_v \quad L_p]$ are calculated according to Eq.(28), and the results are shown in Fig. 2. It can be inferred that gain $K_v$ has the most significant effect on the time delay margin, while $K_P$ and $L_p$ only have apparent influence at large values.

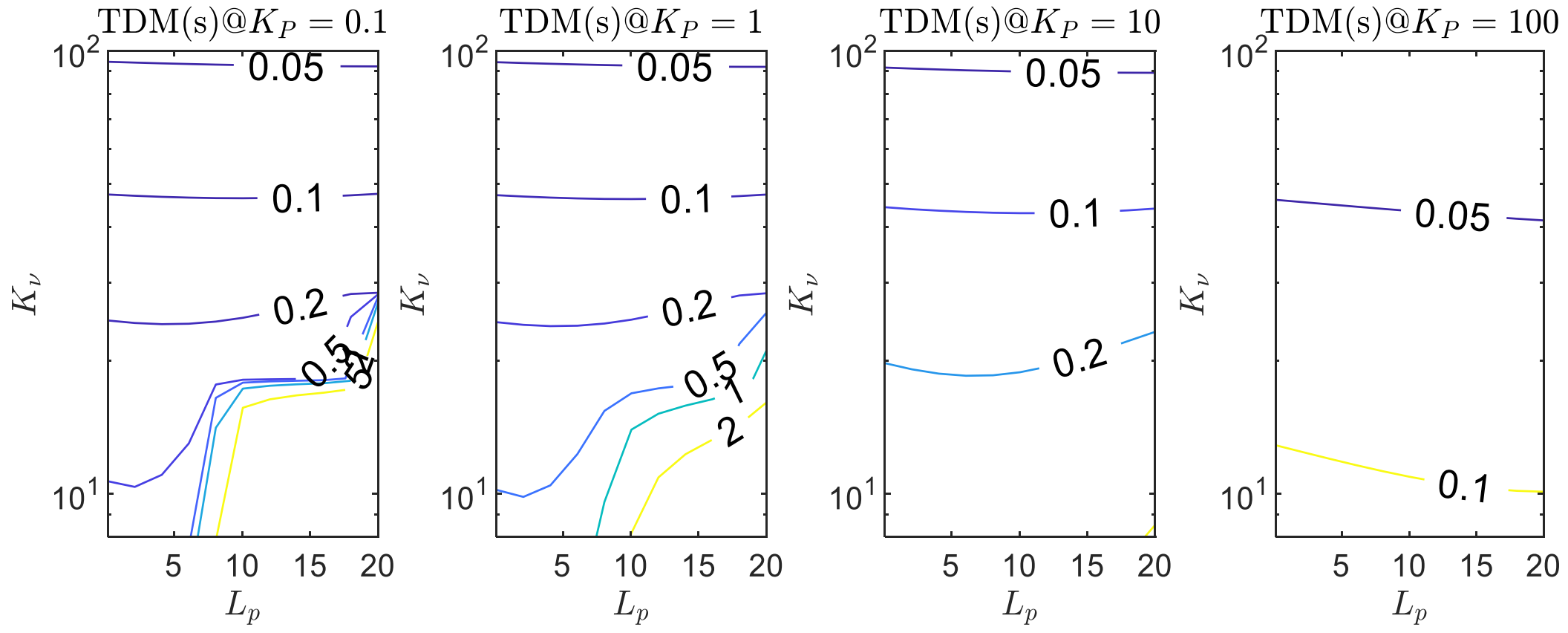

**Fig. 2 Time delay margin of the incremental controller for roll dynamics.**

## 3.2 Time delay

Based on the ideal case described above, we additionally consider the time delays in the actuator, sensor, and input measurement, i.e. $H(s) = e^{-\tau_s s}$, $G_a(s) = \frac{e^{-\tau_a s}}{T_{act}s+1}$, $G_{am}(s) = e^{-\tau_{am}s}$.

The open-loop transfer function and closed-loop transfer function become:

$$L_u = \frac{T_{act}K_v(K_P + s)e^{-(\tau_a+\tau_s)s}}{\hat{B}[T_{act}s + 1 - e^{-(\tau_a+\tau_{am})s}]}P(s) \tag{30}$$

$$T_{yc} = \frac{K_r}{s + K_r} \cdot \frac{T_{act}K_v(K_P + s)P(s)e^{-\tau_a s}}{\hat{B}[T_{act}s + 1 - e^{-(\tau_a+\tau_{am})s}] + T_{act}K_v(K_P + s)P(s)e^{-(\tau_a+\tau_s)s}} \tag{31}$$

Define $\tau_1 = \tau_a + \tau_s, \tau_2 = \tau_a + \tau_{am}$ and use the second-order Padé approximant [18]:

$$e^{-\tau s} \approx \frac{12 - 6\tau s + (\tau s)^2}{12 + 6\tau s + (\tau s)^2} \tag{32}$$

Consider the following two situations:

a) $\tau_1 = \tau_2$, which means the time delays $\tau_s$ and $\tau_{am}$ are well synchronized.
In this case, the open-loop transfer function becomes:

$$L_u = \frac{K_v(K_P + s)(12 - 6\tau_1 s + (\tau_1 s)^2)}{\hat{B}s\left(12 + 12\frac{\tau_1}{T_{act}} + 6\tau_1 s + \tau_1^2 s^2\right)}P(s) \tag{33}$$

Compare Eq.(26) and Eq.(33),the effect of synchronized-delay is equivalent to the transfer function:

$$\Gamma_1(s) = \frac{12 - 6\tau_1 s + (\tau_1 s)^2}{12 + 12\frac{\tau_1}{T_{act}} + 6\tau_1 s + \tau_1^2 s^2} \tag{34}$$

This transfer function $\Gamma_1(s)$ has a similar "phase-lagging" effect to the Padé approximant of the pure delay $e^{-\tau_1 s}$. Besides, it will also reduce the magnitude $|L_u(j\omega)|$ in the low-frequency range by $-20\log(1 + \frac{\tau_1}{T_{act}})$ dB. Therefore, both the phase margin and gain margin of the closed-loop is reduced, as illustrated in Fig. 3, where the red line represents an increased synchronized-delay.

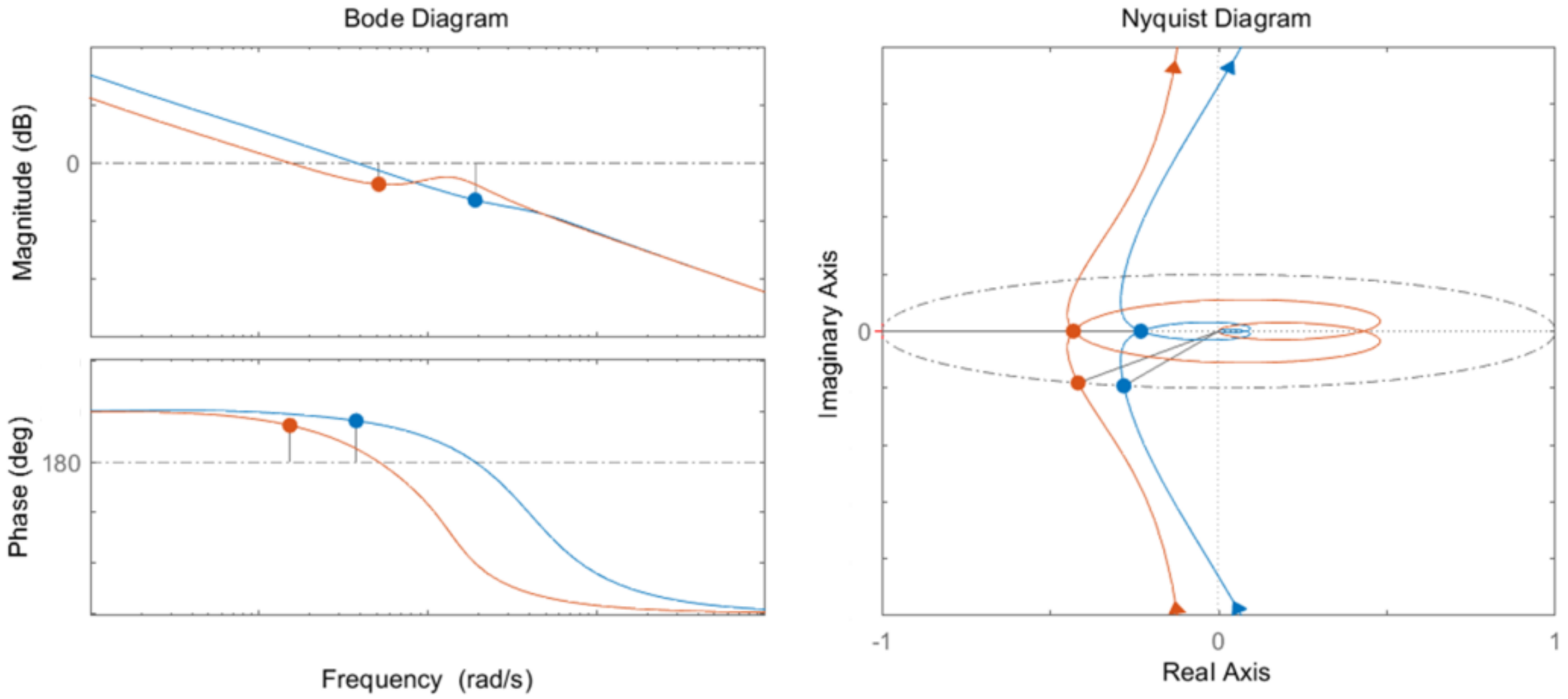

**Fig. 3 Bode diagram and Nyquist diagram of the open-loop with synchronized delay.**

To determine how much synchronized-delay the system can tolerate, the closed-loop transfer function with synchronized-delay is investigated:

$$T_{yc} = \frac{K_r}{s + K_r} \cdot \frac{K_v(K_P + s)e^{-\tau_a s}}{\frac{\hat{B}}{P(s)}\left[s(12 + 6\tau_1 s + \tau_1^2 s^2) + 12\frac{\tau_1}{T_{act}}s\right] + K_v(K_P + s)(12 - 6\tau_1 s + \tau_1^2 s^2)} \tag{35}$$

Assume $\hat{B} = CB$, then $\frac{\hat{B}}{P(s)} = |sI - A|$, and the characteristic polynomial can be obtained as:

$$\det(s) = |sI - A|\left[s(12 + 6\tau_1 s + \tau_1^2 s^2) + 12\frac{\tau_1}{T_{act}}s\right] + K_v(K_P + s)(12 - 6\tau_1 s + \tau_1^2 s^2) \tag{36}$$

Take the roll-mode dynamics in Example 1 for instance. The characteristic polynomial in this case is:

$$\begin{aligned}\det(s) = \tau_1^2 s^4 + \left(6\tau_1 + K_v\tau_1^2 - L_p\tau_1^2\right)s^3 + \left(12\frac{\tau_1}{T_{act}} - 6L_p\tau_1 - 6K_v\tau_1 + K_P K_v\tau_1^2 + 12\right)s^2 \\ + \left(12K_v - 12L_p\left(1 + \frac{\tau_1}{T_{act}}\right) - 6K_P K_v\tau_1\right)s + 12K_P K_v \end{aligned} \tag{37}$$

According to the Routh criterion, the necessary and sufficient condition for closed-loop stability is:

$$\begin{cases} a_i > 0, i = 0,1,2,3,4 \\ \Delta = (a_2 a_3 - a_4 a_1)a_1 - a_3^2 a_0 > 0 \end{cases} \tag{38}$$

Considering that $K_v$ does not exceed $1/T_{act}$ in practice, the solution to the above inequality is:

$$\begin{cases} 0 < \tau_1 < \dfrac{2K_v - 2L_p}{K_P K_v + 2L_p/T_{act}}, & K_P K_v > -2L_p/T_{act} \\ 0 < \tau_1, & K_P K_v < -2L_p/T_{act} \end{cases} \tag{39}$$

Eq.(39) suggests that when the product of control gain $K_P K_v$ is large enough, the system can only tolerate limited synchronized delay.

b) $\tau_1 \neq \tau_2$, which denotes the time delays between $\tau_s$ and $\tau_{am}$ are not synchronized.
In this case, the open-loop transfer function becomes:

$$L_u = \frac{K_v(K_P + s)(12 - 6\tau_1 s + \tau_1^2 s^2)(12 + 6\tau_2 s + \tau_2^2 s^2)}{\hat{B}s(12 + 6\tau_1 s + \tau_1^2 s^2)\left(12 + 12\frac{\tau_2}{T_{act}} + 6\tau_2 s + \tau_2^2 s^2\right)}P(s) \tag{40}$$

Compare Eq.(40) and Eq.(33), it can be inferred that the effect of asynchronized-delay is identical to the transfer function defined as:

$$\Gamma_2(s) = \frac{(12 - 6\tau_1 s + \tau_1^2 s^2)(12 + 6\tau_2 s + \tau_2^2 s^2)}{(12 + 6\tau_1 s + \tau_1^2 s^2)\left(12 + 12\frac{\tau_2}{T_{act}} + 6\tau_2 s + \tau_2^2 s^2\right)} \tag{41}$$

Similar to $\Gamma_1(s)$, transfer function $\Gamma_2(s)$ also has both "phase-lagging" and "low-frequency magnitude reducing" effects while they vary with the relation between $\tau_1$ and $\tau_2$. Set $T_{act}$ to 1/50s, the bode diagram of $\Gamma_2(s)$ with different values of $\tau_1$ and $\tau_2$ is shown in the left figure of Fig.4. And the corresponding open-loop bode diagram for the roll dynamics incremental controller is plotted in the right figure, where $K_v = 1/T_{act}, K_P = 5, L_p = 5, \widehat{L_{\delta_a}} = L_{\delta_a}$. Compared with the case when $\tau_1 = \tau_2$, more phase lag and fewer magnitude reduction will be brought to the open-loop when $\tau_1 > \tau_2$, both of which contribute to a reduced gain margin. On the other hand, when $\tau_1 < \tau_2$, $\omega_{\angle L_u=180°}$ increase, but the magnitude at this frequency increase, so the gain margin is reduced again as a result.

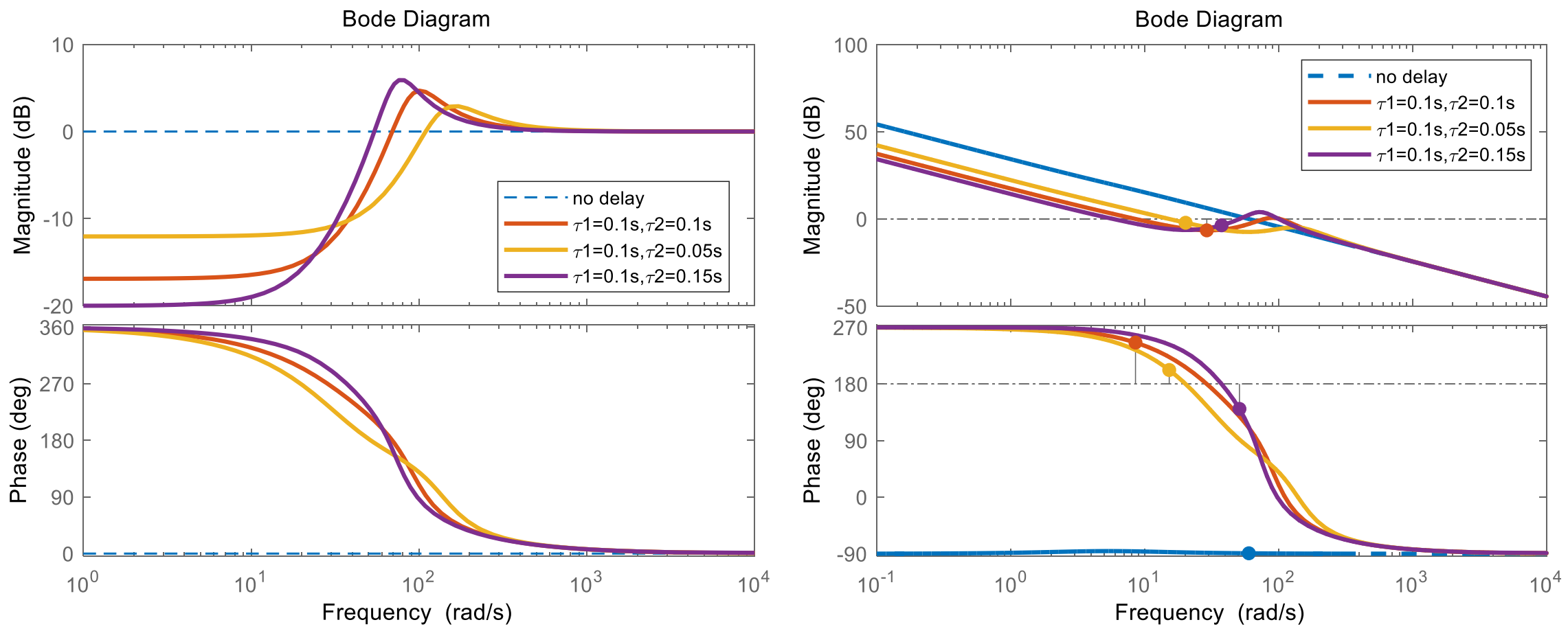

**Fig. 4  Bode diagram of $\Gamma_2$ and the open-loop with asynchronized delays.**

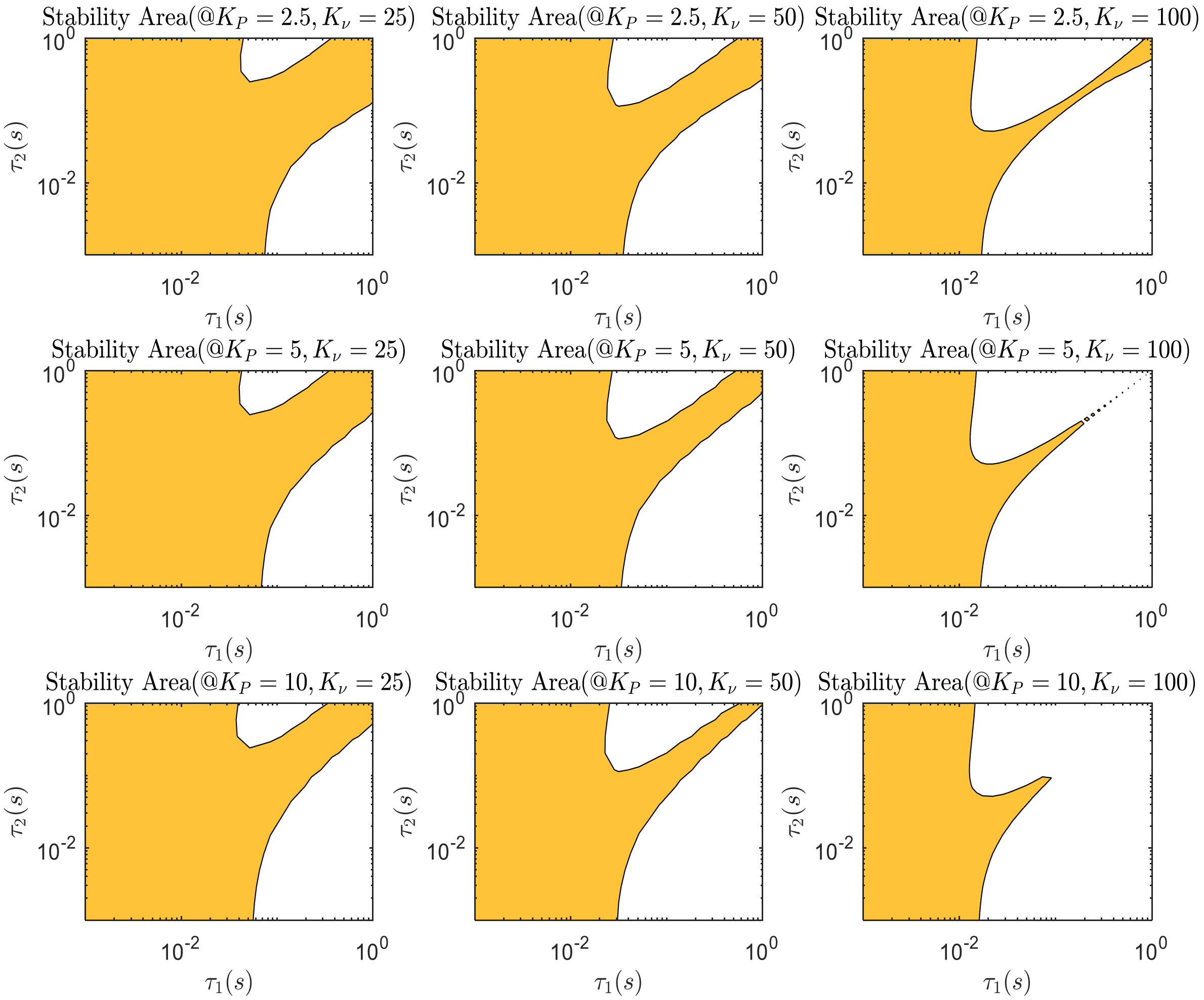

**Fig. 5  Stability area of $[\tau_1, \tau_2]$ for the incremental controller with different gains.**

A detailed illustration of the relationship between stability and asynchronized-delay $[\tau_1, \tau_2]$ is demonstrated in Fig.5, where the colored zone represents stability. Three findings can be drawn from Fig.5: Firstly, closed-loop stability is independent of $\tau_2$ as long as $\tau_1$ is small enough, which means actuator measurements delay will not drive the closed-loop unstable when sensor delay is below the delay margin. Secondly, stability could be maintained with a considerable value of $\tau_1$ under the condition that $\tau_2$ is close enough to $\tau_1$, that is, delay synchronization. Thirdly, larger control gain $K_P$ and $K_v$ will contract the stability area. Significantly, they will impair the capability of delay synchronization.

Combining the analysis results of a) and b), a conclusion can be drawn that the incremental controller is very sensitive to sensor time delay, but the tolerance to delay could be enhanced by introducing a synchronization delay in the actuator measurement loop. However, as the synchronized delays increase altogether, the stability margin will decrease, and the benefits of synchronization will be infringed by high control gains.

## 3.3 Sensor and filter dynamics compensation

The sensor and filter dynamics themselves also have a “time delay” effect. This part investigates the impact of compensation for the sensor and filter dynamics into the actuator measurement loop.

When neither sensor nor filter dynamics compensation is incorporated in the actuator measurement loop, the open-loop transfer function is given by:

$$L_{u1} = \frac{K_v\left(K_P + s + K_P T_{diff} s\right)}{\hat{B} s\left(T_{diff} s + 1\right)\left(T_{sensor} s + 1\right)} P(s) \tag{42}$$

When filter dynamics compensation alone is incorporated in the actuator measurement loop, one obtains:

$$L_{u2} = \frac{K_v\left(K_P + s + K_P T_{diff} s\right)}{\hat{B} s\left(T_{diff} s + 1 + \frac{T_{diff}}{T_{act}}\right)\left(T_{sensor} s + 1\right)} P(s) \tag{43}$$

When both the sensor and filter dynamics compensation are introduced in the actuator measurement loop, the open-loop transfer function is:

$$L_{u3} = \frac{K_v\left(K_P + s + K_P T_{diff} s\right)}{\hat{B} s\left[\left(T_{diff} s + 1 + \frac{T_{diff}}{T_{act}}\right)\left(T_{sensor} s + 1\right) + \frac{T_{sensor}}{T_{act}}\right]} P(s) \tag{44}$$

Therefore, the relation between these three loop-transfers can be obtained:

$$\begin{gathered} |L_{u1}(j\omega)| > |L_{u2}(j\omega)| > |L_{u3}(j\omega)| \\ \angle L_{u3}(j\omega) > \angle L_{u2}(j\omega) > \angle L_{u1}(j\omega) \end{gathered} \tag{45}$$

It suggests that incorporating sensor and filter dynamics compensation into the actuator measurement loop will increase the gain margin, phase margin, and time delay margin. Based on the above analysis, we can infer that introducing compensation for any additional linear dynamics aimed at measurements noise attenuation into the $G_{am}$ loop will improve stability.

## 3.4 Pseudo-control-hedging

Although pseudo-control-hedging was originally used to cope with the nonlinear characteristics of the actuator like saturation [14], only its linear dynamics are considered here. The open-loop functions with PCH and without PCH are Eq.(21) and Eq.(23) respectively. The ration function between these two is:

$$R(s) = \frac{L_{u,PCH}}{L_u} = \frac{s + K_r}{s + K_r + K_v T_{act}(K_P - K_r)} \tag{46}$$

We can infer that in the case of $K_P > K_r$, PCH will reduce the gain and increase the open-loop phase in the low-frequency band; As a result, the stability margins increase. And the effect of PCH will be more prominent with a higher value of $K_v T_{act}$. Conversely, when $K_P < K_r$, PCH will reduce stability, and this reduction will be more severe with a higher value of $K_v T_{act}$. However, the actual situation for flight control is generally the former.

# 4 Performance Analysis for Incremental Controller

This section derives the closed-loop transfer function that characterizes the closed-loop system's performance, including command tracking, robustness, disturbance suppression, and noise suppression, and connects them with the open-loop transfer function, and analyzes the relationship between them.

## 4.1 Tracking and Robustness

When PCH is off, the closed-loop transfer function from command to output measurement is:

$$\begin{aligned} T_{y_m c} &= \frac{K_r T_{act} K_v G_a(s) H(s) P(s)(K_P + s)}{(s + K_r)\left[\hat{B}(1 - G_a(s) G_{am}(s)) + T_{act} K_v G_a(s) H(s) P(s)(K_P + sF(s))\right]} \\ &= \frac{K_r}{s + K_r} \cdot \frac{K_P + s}{K_P + sF(s)} \cdot \frac{L_u}{1 + L_u} \end{aligned} \tag{47}$$

And the tracking error function will be:

$$T_{ec} = T_{rc} - T_{y_m c} = \frac{K_r}{s + K_r}\left[1 - \frac{K_P + s}{K_P + sF(s)} \cdot \frac{L_u}{1 + L_u}\right] \tag{48}$$

Consider an additive uncertainty in the plant dynamics $\Delta P(s)$, then the deviation of command tracking performance due to $\Delta P(s)$ can be given by:

$$\begin{aligned} \Delta T_{yc} = &\frac{K_r T_{act} K_v G_a(s)[P(s) + \Delta P(s)](K_P + s)}{(s + K_r)\left[\hat{B}(1 - G_a(s) G_{am}(s)) + T_{act} K_v G_a(s) H(s)[P(s) + \Delta P(s)](K_P + sF(s))\right]} \\ &- \frac{K_r T_{act} K_v G_a(s) P(s)(K_P + s)}{(s + K_r)\left[\hat{B}(1 - G_a(s) G_{am}(s)) + T_{act} K_v G_a(s) H(s) P(s)(K_P + sF(s))\right]} \end{aligned} \tag{49}$$

The sensitivity function can be obtained as follows:

$$S(s) = \lim_{\Delta P \to 0} \frac{\Delta T_{yc}/T_{yc}}{\Delta P/P} = \frac{\hat{B}\left(1 - G_a(s) G_{am}(s)\right)}{\hat{B}\left(1 - G_a(s) G_{am}(s)\right) + T_{act} K_v G_a(s) H(s) P(s)\left(K_P + sF(s)\right)} \tag{50}$$

Combing Eq.(21) with Eq.(50), the sensitivity function can be written as:

$$S(s) = \frac{1}{1 + L_u} \tag{51}$$

Similarly, when PCH is on

$$T_{ec,PCH} = \frac{K_r}{s + K_r}\left[1 - \frac{K_P + s}{K_P + sF(s)} \cdot \frac{L_{u,PCH}}{1 + L_{u,PCH}}\right] \tag{52}$$

$$S_{PCH}(s) = \frac{1}{1 + L_{u,PCH}} \tag{53}$$

It can be inferred that tracking performance and robustness are consistent, and high loop gain will result in enhanced tracking ability and robustness. According to Eq.(46), when $K_P > K_r$, PCH will undermine tracking ability and robustness, and vice versa.

## 4.2 Disturbance rejection and noise attenuation

When PCH is off, the closed-loop transfer function from disturbance to output in Fig. 1 goes:

$$T_{yd} = \frac{P(s)}{1 + G_a(s)\bar{C}(s)H(s)P(s)} = \frac{P(s)}{1 + L_u} = P(s)S(s) \tag{54}$$

And the closed-loop transfer function from noise to output is:

$$T_{yn} = \frac{-H(s)\bar{C}(s)G_a(s)P(s)}{1 + G_a(s)\bar{C}(s)H(s)P(s)} = \frac{1}{1 + L_u} - 1 = S(s) - 1 \tag{55}$$

When PCH is on, the disturbance and noise functions can be derived similarly:

$$T_{yd,PCH} = \frac{P(s)}{1 + L_{u,PCH}} = P(s)S_{PCH}(s) \tag{56}$$

$$T_{yn,PCH} = \frac{1}{1 + L_{u,PCH}} - 1 = S_{PCH}(s) - 1 \tag{57}$$

Therefore, for a given plant, the expectation of improving its disturbance rejection ability is consistent with the enhancement of robustness but contradicts its noise attenuation. According to Eq.(46), when $K_P > K_r$, PCH will enhance noise attenuation but undermine disturbance rejection, vice versa.

# 5 Numerical Evaluation

In this section, numerical evaluations are conducted for the incremental controller of a short-period model. Comprehensive evaluation metrics are developed to quantitatively evaluate the stability, tracking performance, disturbance rejection ability, noise attenuation ability, and robustness to model uncertainty of the incremental controller. The influence of each element of the control system is thoroughly investigated.

## 5.1 Incremental Control for the Short-Period Model

The short-period model with the wind effect is given as follows [19]:

$$\begin{bmatrix}\dot{\alpha}\\ \dot{q}\end{bmatrix}=\begin{bmatrix}Z_\alpha & Z_q+1\\ M_\alpha & M_q\end{bmatrix}\begin{bmatrix}\alpha\\ q\end{bmatrix}+\begin{bmatrix}Z_\eta\\ M_\eta\end{bmatrix}\eta+\begin{bmatrix}-Z_V & -\frac{Z_\alpha}{V_0}\\ -M_V & -\frac{M_\alpha}{V_0}\end{bmatrix}\begin{bmatrix}u_g\\ w_g\end{bmatrix}$$

The control derivatives of the DA-42 aircraft at the flight condition of $V_0$=70m/s are used and listed in Table. 1. These derivatives are taken as nominal values, a ±30% deviation of each parameter will be considered when performing robustness verifications.

**Table. 1 Parameter of the short-period model of DA-42 at $V_0$=70m/s[20][21]**

| Parameter | $Z_\alpha$ | $Z_q$ | $Z_\eta$ | $Z_V$ | $M_\alpha$ | $M_q$ | $M_\eta$ | $M_V$ |
|---|---|---|---|---|---|---|---|---|
| Value | -1.27 | 0.0037 | 4.4e-4 | -0.003 | -17.71 | -2.63 | -8.18 | 2.8e-04 |

The incremental pitch rate control system is illustrated in Fig. 6. It should be noted that because the PCH modifies the state of the reference model, as well as the error dynamics. Therefore, the absolute error value is obtained by subtracting the system output from the unmodified reference signal.

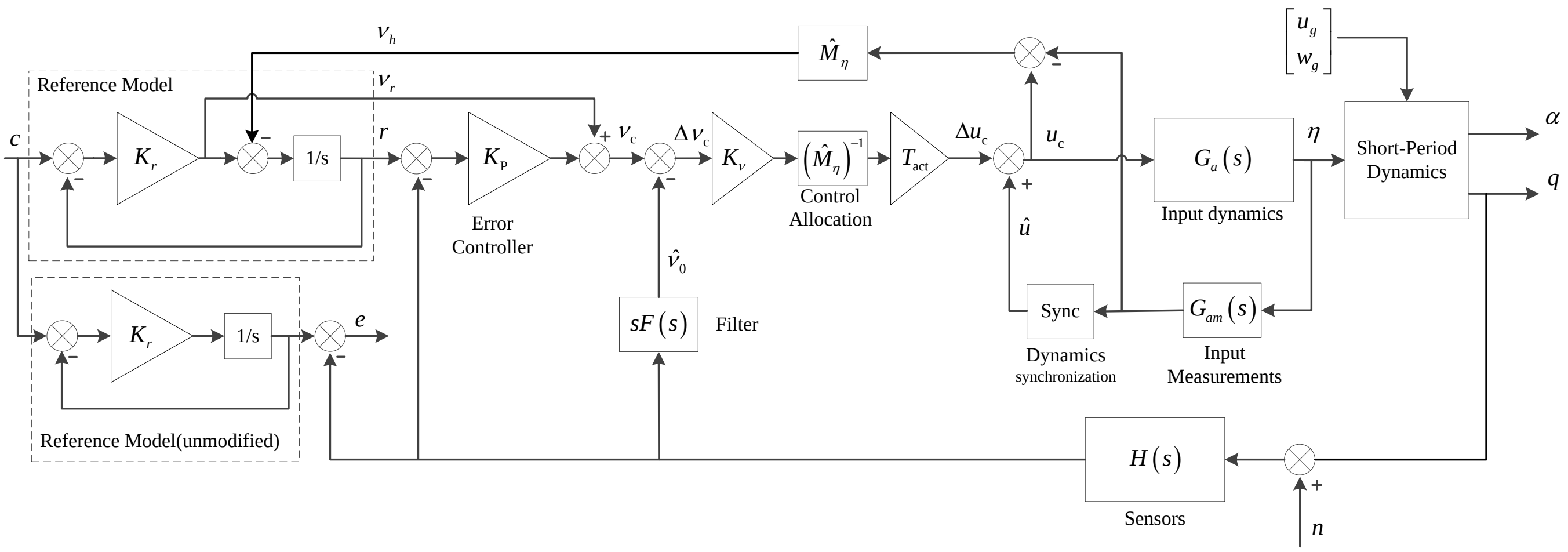

**Fig. 6 Block diagram of the incremental pitch rate control system**

The descriptions of each part of the INDI control system are shown in Table. 2

**Table. 2 Descriptions of the incremental pitch rate control system**

| Elements | Description | Nominal values |
|---|---|---|
| Reference Model | $\nu_r=K_r(c-r),\dot{r}=\nu_r-\nu_h$ | $K_r=5$ |
| Actuator | $\dot{\eta}(t)=sat_{\dot{\eta}}\left\{\frac{u(t-\tau_a)}{T_{\text{act}}}-\frac{sat_\eta[\eta(t)]}{T_{\text{act}}}\right\}$ | $T_{\text{act}}=1/60(s),\tau_a=0$<br>$sat_{\dot{\eta}}=\{\dot{\eta}\|\|\dot{\eta}\|<100°/s\}$<br>$sat_\eta=\{\eta\|\|\eta\|<30°\}$ |
| Sensor | $\dot{q_m}(t)=\frac{q(t-\tau_s)}{T_{\text{sensor}}}-\frac{q_m(t)}{T_{\text{sensor}}},$ | $T_{\text{sensor}}=1/300(s),\tau_s=0$<br>Noise $(\sigma^2)=4.0\times10^{-7}$ |
| Filter | $\dot{q}_f = \frac{s}{T_{\text{diff}}s+1} q_m$ | $T_{\text{diff}} = 1/30\ (s)$ |
| Actuator measurements | $G_{am}(s) = \frac{e^{-\tau_{am}s}}{(T_{\text{diff}}s+1)(T_{\text{sensor}}s+1)}$ | $\tau_{am} = 0$ |
| Error gain | $\dot{q}_c = K_P(q_r - q_m)$ | $K_P = 8$ |
| Control effectiveness estimate | $\widehat{M}_\eta$ | $\widehat{M}_\eta = M_\eta$ |
| Disturbance | $u_g = \frac{u_m}{2}\left(1 - cos\left(\frac{\pi x}{d_x}\right)\right) \quad 0 \le x \le d_x$<br>$w_g = \frac{w_m}{2}\left(1 - cos\left(\frac{\pi x}{d_z}\right)\right) \quad 0 \le x \le d_z$ | $d_x = 120m$，$d_z = 80m$，<br>$u_m = 3.5m/s$，$w_m = 3m/s$ |

## 5.2 Evaluation Metrics

The overall performance of the incremental controller is examined through a series of simulations, and evaluation metrics are extracted in the meantime, as shown in Fig. 7. For command tracking evaluation, the $q$ command is designed as a two-way square wave signal with an amplitude of 10°/s and an interval of 3s, and the root-mean-square (RMS) tracking error and input are recorded. For disturbance rejection evaluation, both horizontal and vertical gust-winds are activated at t=3s. For noise attenuation evaluation, white noise with variance $\sigma^2$=4.0×$10^{-7}$ is added to the sensor measurements, and the RMS of tracking error and input are recorded. For robustness verification, command tracking simulations are done repeatedly for N=100 samples of the uncertain plant described above. The standard deviation of all N=100 RMS tracking errors is calculated to characterize the performance deviation due to model uncertainties. The descriptions of all the evaluation metrics and their values at the nominal condition are shown in Table. 3.

**Table. 3 Evaluation metrics for the incremental pitch rate control system**

| Category | Evaluation Metric | Description | Nominal value |
|---|---|---|---|
| Stability | GM | Gain Margin | 21.89dB |
| | PM | Phase Margin | 62.74° |
| | TDM | Time Delay Margin | 0.0485s |
| Command tracking | RMSer | RMS error for tracking | 0.9014°/s |
| | RMSur | RMS input for tracking | 14.3805° |
| Disturbance rejection | RMSed | RMS error due to disturbance | 0.0773°/s |
| | RMSud | RMS input due to disturbance | 1.85° |
| Noise attenuation | RMSen | RMS error due to noise | 0.0325°/s |
| | RMSun | RMS input due to noise | 0.0641° |
| Robustness | $\sigma(\text{RMSer}_i), i = 1,2,\dots,N$ | Standard deviation of RMS error | 0.115°/s |

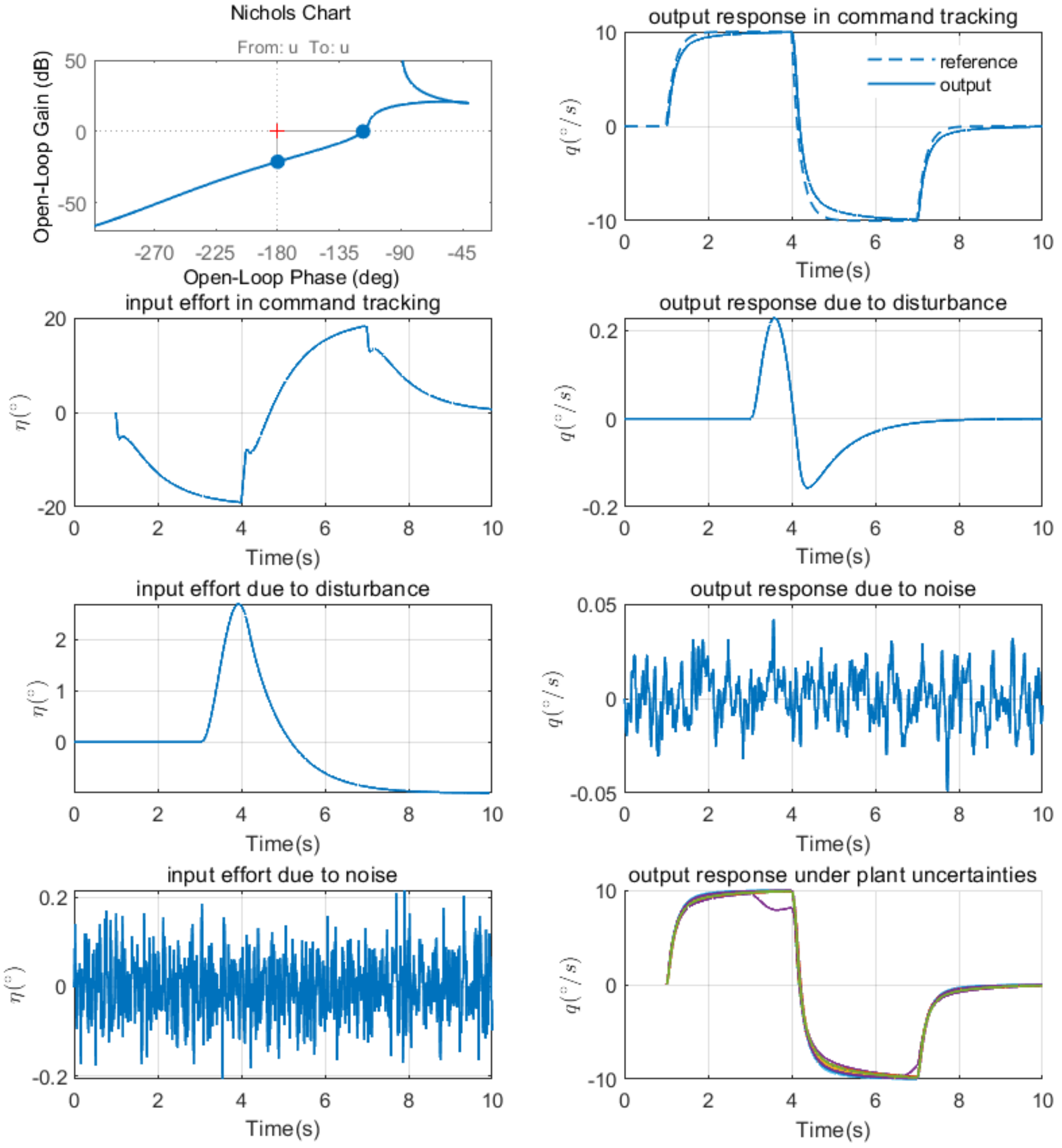

**Fig. 7 Simulation results of the incremental control system under the nominal condition**

## 5.3 Results and Analysis

The evaluation results are shown in Fig. 8. Each subplot represents the effect of one or more specific parameters on the whole incremental controller. Here is a brief analysis of each:

It can be learned from Fig. 8(a) that a larger gain $K_P$ will improve the command tracking, robustness, and disturbance rejection at the cost of stability and noise attenuation. Besides, when $K_P < K_r$, PCH will improve tracking and disturbance rejection while hurt stability and noise attenuation, and vice versa when $K_P > K_r$. Fig. 8(b) suggests that gain $K_v$ has similar effects as $K_P$ while it is much more essential to stability margins. Fig. 8(c) reveals that it is more favorable to appropriately over-estimate control effectiveness than under-estimate it. Fig. 8(d) indicates that higher actuator frequencies lead to better tracking performance, disturbance rejection, and robustness, while decreased stability and noise attenuation. Besides, PCH will improve stability for slower actuators. Fig. 8(e) and Fig. 8(f) show that the introduction of sensor and filter dynamics compensation into the actuator measurement loop can significantly enhance system stability and improve noise attenuation. Still, it will adversely affect command tracking, disturbance suppression, and robustness when the sensor frequency or filter bandwidth is low. At last, Fig. 8(g) demonstrates that the delay-synchronized system has enhanced stability and improved performances.

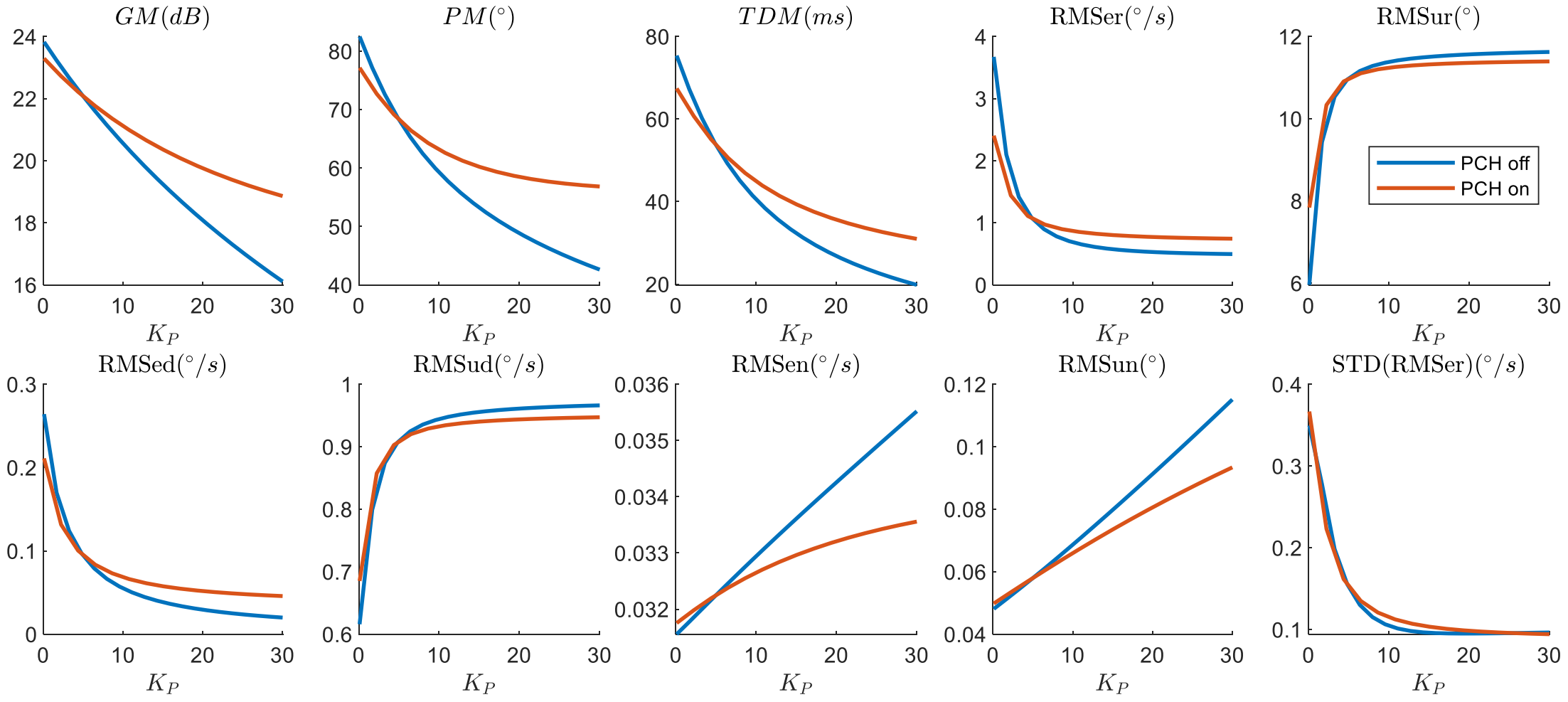

a) Influence of error gain $K_P$ on evaluation metrics

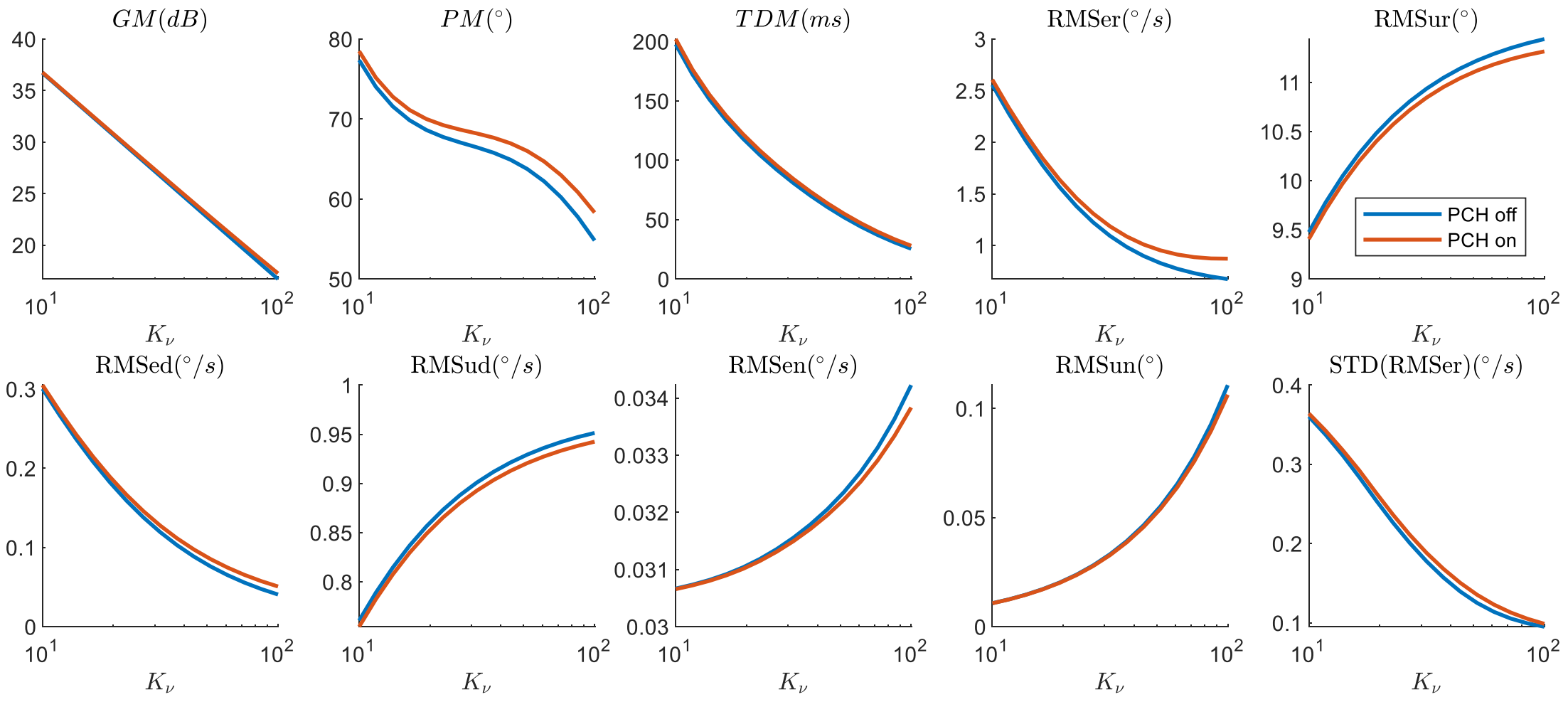

b) Influence of gain $K_\nu$ on evaluation metrics

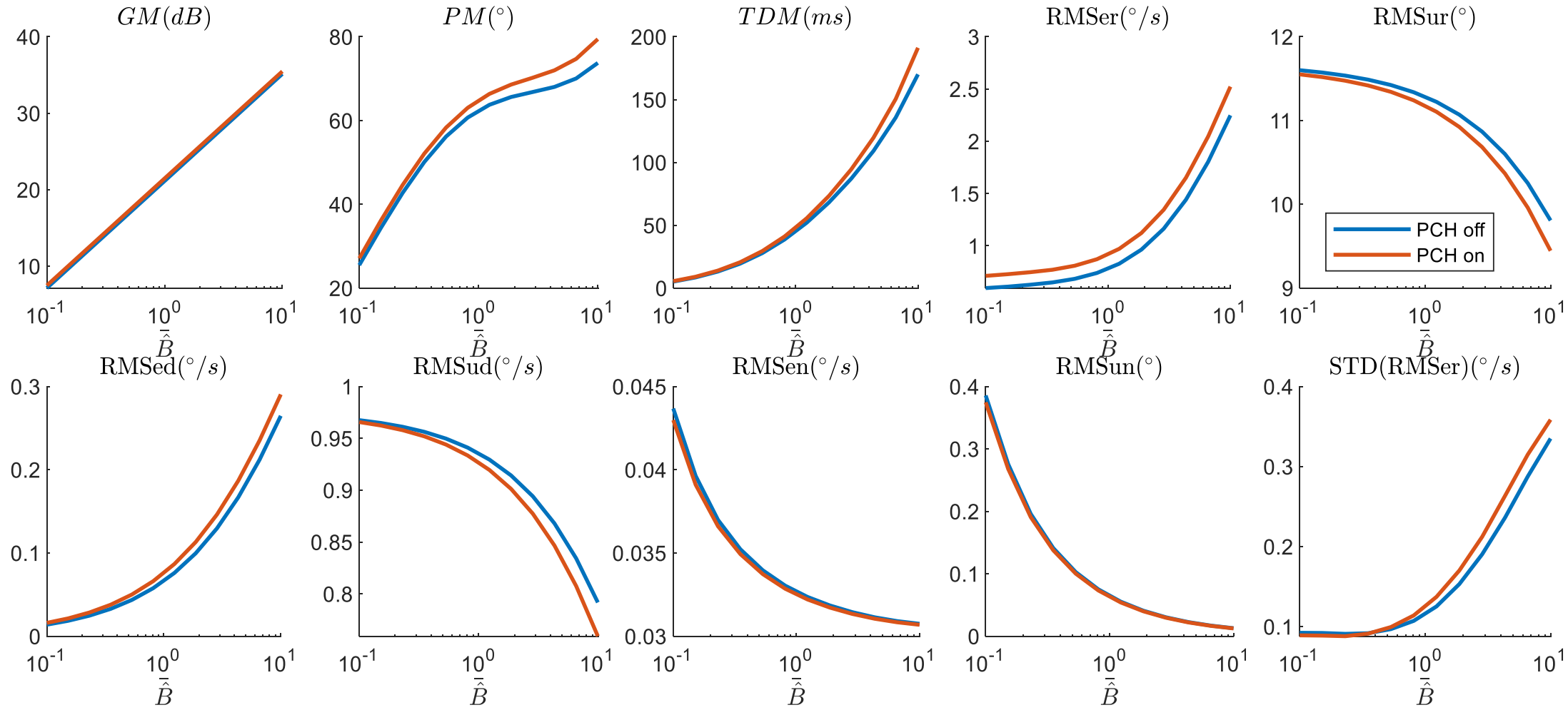

c) Influence of control effectiveness scaling on evaluation metrics

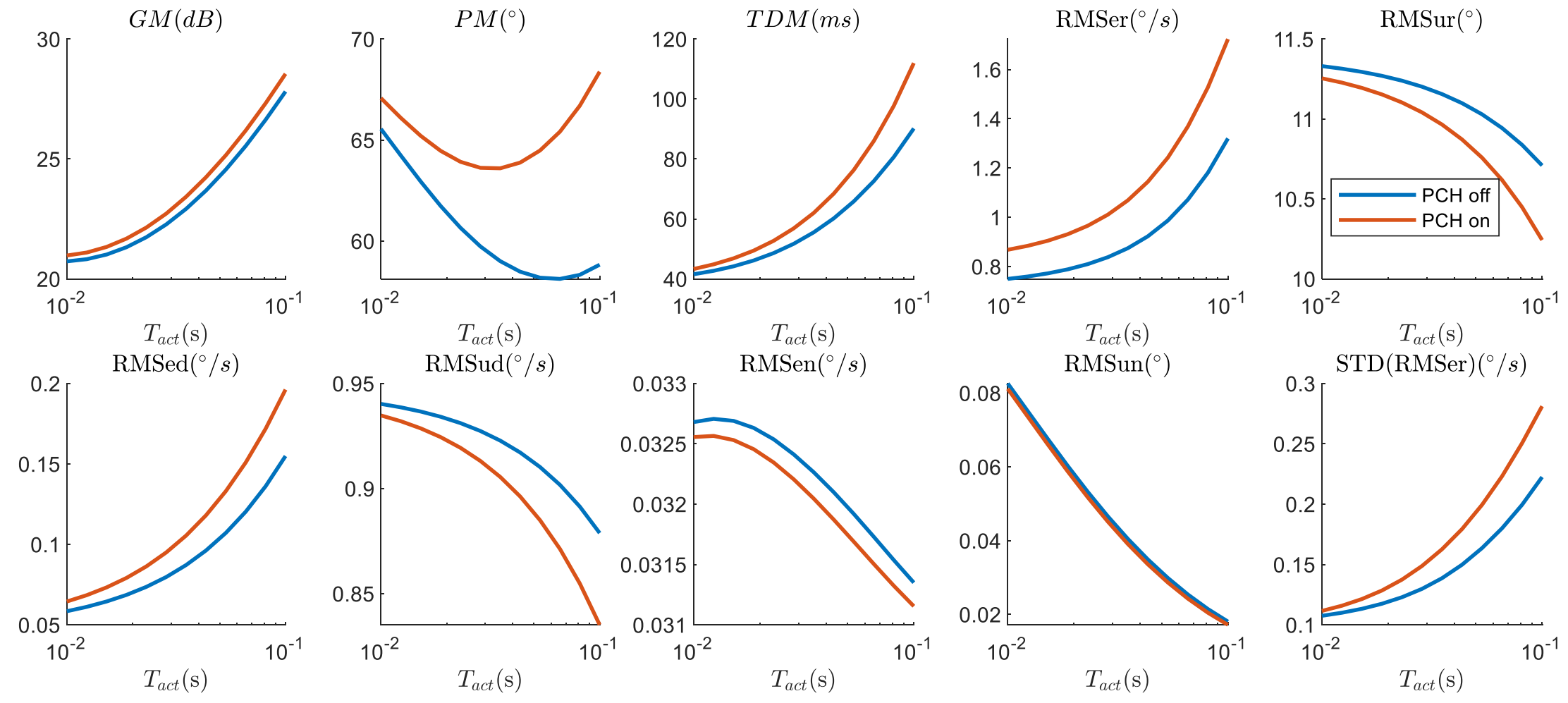

d) Influence of actuator frequency evaluation metrics

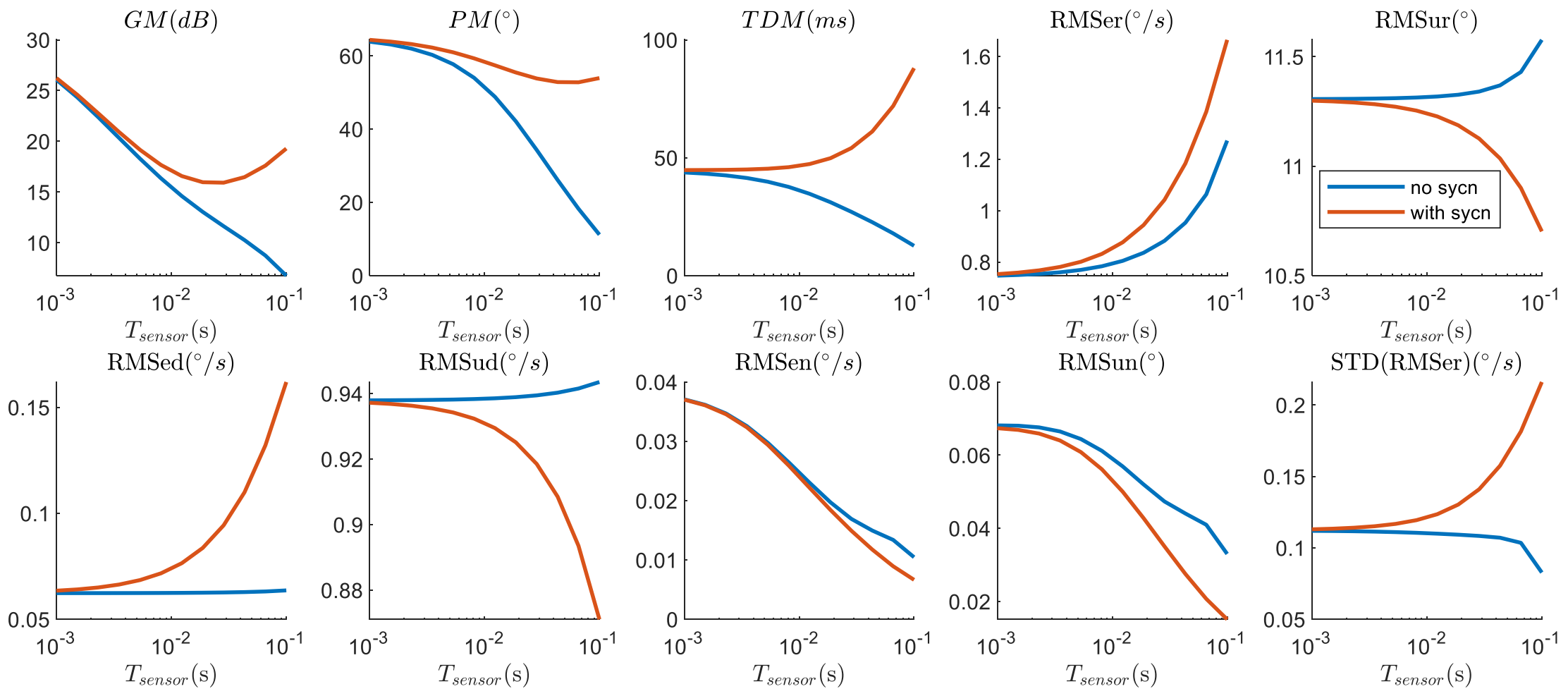

e) Influence of sensor frequency on evaluation metrics

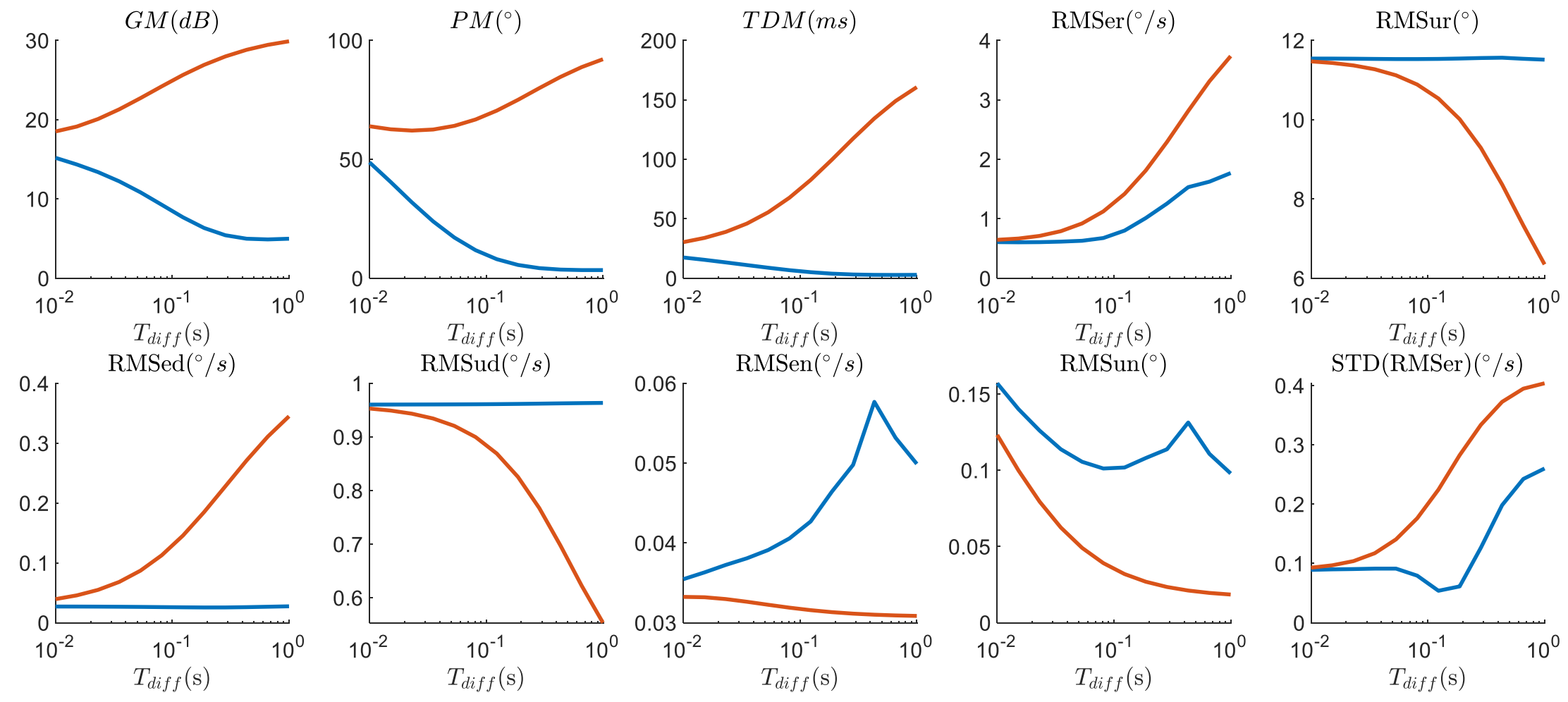

f) Influence of filter bandwidth on evaluation metrics

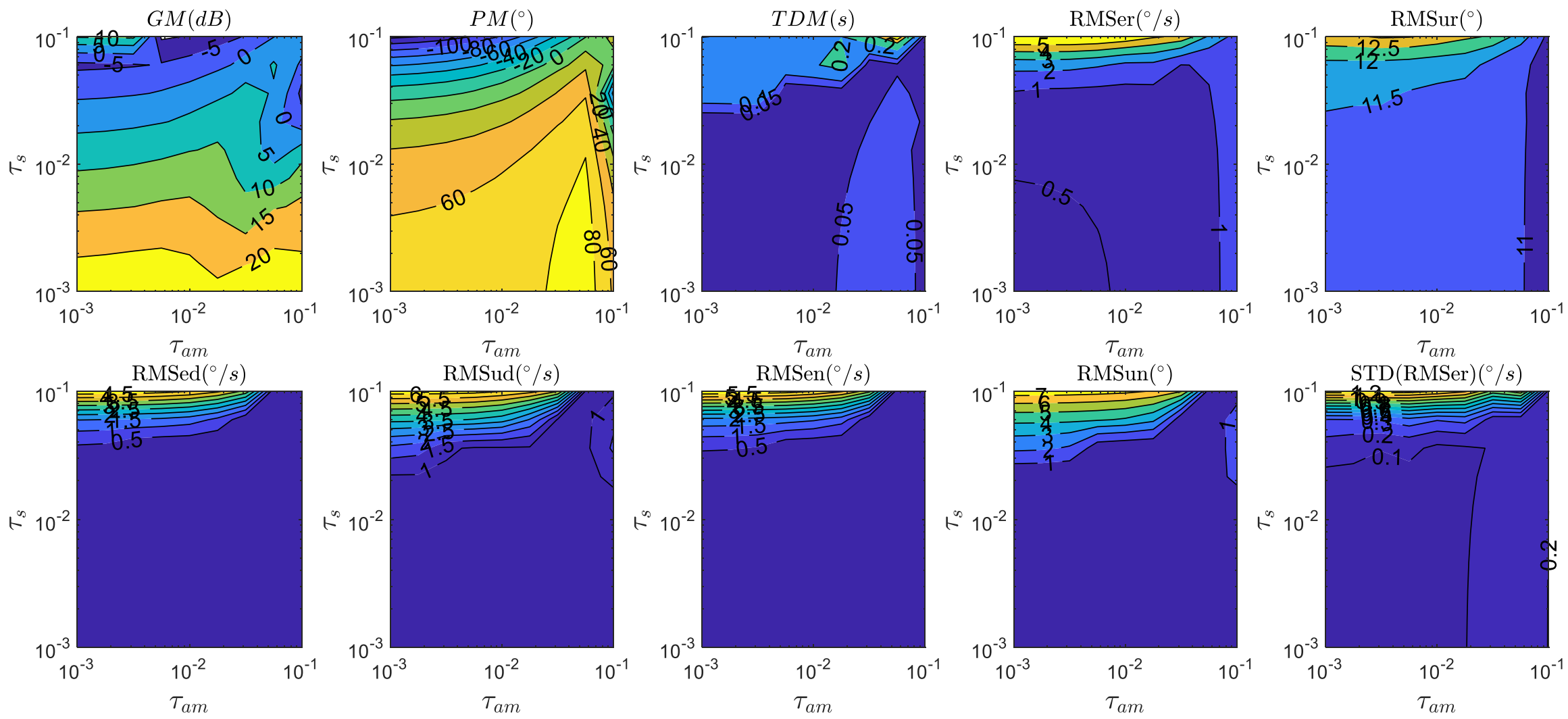

g) Influence of sensor delay and input measurement delay on evaluation metrics

**Fig. 8 Influence of different elements of the incremental control system on its overall performance**

# 6 Conclusion

This paper has discussed the stability and performance of SISO linear incremental flight control systems in both analytical and numerical methods. The following conclusions can be drawn:

1) The advantages of incremental control are better tracking performance, robustness, and disturbance rejection ability, which are mainly attributed to high loop gain caused by the direct correspondence between the pseudo control and the control input. On the other hand, the major challenges are reduced stability margins and insufficient noise attenuation. The modified incremental controller has extra design freedom, the pseudo control gain, which helps to establish a balance between the above two aspects.
2) Delay synchronization between sensor measurements and actuator measurements can extend stability. Besides transport delay, linear dynamics of sensor and filter should also be compensated for in the actuator measurement loop.
3) PCH incorporates the reference model into the closed-loop. When the error control gain is greater than the reference model gain, PCH will increase stability margins and improve the noise attenuation ability, but weaken the tracking and disturbance rejection, and the benefits brought by PCH is more evident under the conditions of high pseudo control gain and slow actuator dynamics. The contrary is the case when the error control gain is less than the reference model gain.
4) Higher error gain, smaller control effectiveness estimate, higher actuator frequency, higher sensor frequency, and higher filter bandwidth are beneficial to command tracking, robustness, and disturbance rejection. At the same time, they may hurt stability and noise suppression.

This work provides some insights on the INDI flight control in a practical context. Further work should be done for more complex dynamic systems such as MIMO and nonlinear flight systems.

# References

[1] Balas, G. J. Flight Control Law Design: An Industry Perspective. *European Journal of Control*, 9(2-3), 2003.

[2] Enns, D., Bugajski, D., Hendrick, R., and Stein, G. Dynamic Inversion: An Evolving Methodology for Flight Control Design. *International Journal of Control*, 59(1), 1994. DOI: 10.1080/00207179408923070.

[3] Smith, P. A Simplified Approach to Nonlinear Dynamic Inversion Based Flight Control. *23rd Atmospheric Flight Mechanics Conference*, American Institute of Aeronautics and Astronautics, Reston, VA, 1998.

[4] Smeur, E. J., de Croon, G. C., and Chu, Q. Gust Disturbance Alleviation with Incremental Nonlinear Dynamic Inversion. *2016 IEEE/RSJ International Conference on Intelligent Robots and Systems (IROS)*, IEEE, 2016. DOI: 10.1109/IROS.2016.7759827.

[5] Falconí G P, Schatz S P, Holzapfel F. Fault tolerant control of a hexarotor using a command governor augmentation. *2016 24th Mediterranean Conference on Control and Automation (MED)*. IEEE, 2016. DOI: 10.1109/MED.2016.7535981.

[6] Akkinapalli, Venkata Sravan, and Florian Holzapfel. "Incremental Dynamic Inversion based Velocity Tracking Controller for a Multicopter System." *2018 AIAA Guidance, Navigation, and Control Conference*. Kissimmee, Florida, January 2018. DOI:10.2514/6.2018-1345.

[7] Wang X, Van Kampen E J, Chu Q, et al. Stability analysis for incremental nonlinear dynamic inversion control. *Journal of Guidance, Control, and Dynamics*, 42(5), 2019. DOI: 10.2514/1.G003791.

[8] Stéphane Delannoy and Simon Oudin. Longitudinal control law for modern long-range civil aircrafts. In *Proceedings of the 2nd CEAS Specialist Conference on Guidance, Navigation & Control (EuroGNC 2013),* Delft, Netherland, 2013.

[9] Sieberling, S., Q. P. Chu, and J. A. Mulder. "Robust flight control using incremental nonlinear dynamic inversion and angular acceleration prediction." Journal of guidance, control, and dynamics 33.6 (2010): 1732-1742. DOI:10.2514/1.49978

[10] Raab S A, Zhang J, Bhardwaj P, et al. Consideration of Control Effector Dynamics and Saturations in an Extended INDI Approach. *AIAA Aviation 2019 Forum*. Dallas, USA, 2019. DOI: 10.2514/6.2019-3267.

[11] Koschorke, J., Falkena, W., Van Kampen, E.-J., & Chu, Q. P.. Time Delayed Incremental Nonlinear Control. *AIAA Guidance, Navigation, and Control (GNC) Conference*. Boston, USA, 2013. DOI:10.2514/6.2013-4929.

[12] Smith, P. & Berry, A.. Flight Test Experience of a Non-Linear Dynamic Inversion Control Law on the VAAC Harrier. In *Atmospheric Flight Mechanics Conference*. Denver, USA, 2000. DOI:10.2514/6.2000-3914.

[13] Jeon B J, Seo M G, Shin H S, et al. Understandings of the Incremental Backstepping Control through Theoretical Analysis under the Model Uncertainties. *2018 IEEE Conference on Control Technology and Applications (CCTA)*. IEEE, Copenhagen, Denmark, 2018. DOI: 10.1109/CCTA.2018.8511479.

[14] Johnson, Eric N., and Anthony J. Calise. Pseudo-control hedging: A new method for adaptive control. *Advances in navigation guidance and control technology workshop*. Alabama, USA, 2000. DOI: 10.2514/6.2018-0845.

[15] Raab, Stefan A., et al. Proposal of a unified control strategy for vertical take-off and landing transition aircraft configurations. *2018 Applied Aerodynamics Conference*. Atlanta, USA, 2018. DOI: 10.2514/6.2018-3478.

[16] Keijzer, Twan, et al. Design and flight testing of incremental backstepping based control laws with angular accelerometer feedback. *AIAA Scitech 2019 Forum*. San Diego, USA, 2019. DOI: 10.2514/6.2019-0129.

[17] Cook, Michael V. *Flight dynamics principles: a linear systems approach to aircraft stability and control*. Butterworth-Heinemann, 2012, pp. 198–199.

[18] Vajta, Miklos. Some remarks on Padé-approximations. In Proceedings of the 3rd TEMPUS-INTCOM Symposium. Veszprém, Hungary, 2000.

[19] Stevens, Brian L., Frank L. Lewis, and Eric N. Johnson. Aircraft control and simulation: dynamics, controls design, and autonomous systems. John Wiley & Sons, 2015.

[20] Karlsson, Erik, et al. "Automatic flight path control of an experimental DA42 general aviation aircraft." 2016 14th International Conference on Control, Automation, Robotics and Vision (ICARCV). IEEE, 2016. DOI:10.1109/icarcv.2016.7838566

[21] Krause, Christoph, and Florian Holzapfel. "System Automation of a DA42 General Aviation Aircraft." 2018 Aviation Technology, Integration, and Operations Conference. 2018. DOI: 10.2514/6.2018-3984